\documentstyle[preprint,prl,aps]{revtex}
\begin{document}
\draft
\title{Critical Behavior of an Interacting Monomer--dimer Model.}
\author{Mann Ho Kim and Hyunggyu Park}
\address{Department of physics, Inha university, Inchon, 402-751, Korea}
\maketitle
\begin{abstract}
We study a monomer-dimer model with repulsive
interactions between the same species in one dimension.
With infinitely strong interactions the model exhibits a continuous
transition from a reactive phase to an inactive phase with two equivalent
absorbing states. Monte Carlo simulations show that the critical
behavior is different from the conventional directed percolation
universality class but seems to be consistent with that of the models
with the mass conservation of modulo 2.
\end{abstract}

A monomer-dimer model was introduced by Ziff, Gulari, and Barshad
to describe the oxidation of carbon monoxide on catalytic surface
\cite{ZGB}.
In this model, a monomer ($CO$) adsorbs onto a single vacant site, while
a dimer ($O_2$) adsorbs onto a pair of adjacent vacant sites and then
immediately
dissociates. A nearest neighbor of adsorbates, comprised of a dissociated
$O$ atom and an $CO$ atom, reacts and forms a $CO_2$ molecule and desorbs
from the metal surface.
In two dimensions, as the $CO$ gas pressure is lowered, the system
undergoes a first-order
transition from a $CO$-saturated inactive phase into a reactive
steady state and then a continuous transition into a $O_2$-saturated
inactive phase. This continuous transition
is shown to be
in the same universality class as the directed percolation (DP) and the
Reggeon field theory (RFT).

Motivated by the monomer-dimer model,
there have appeared many related lattice models
to study nonequilibrium phase transitions.
One of the simplest models is the monomer-monomer
model\cite{Meak87,Ziff}
in which particles of two different species can adsorb on a single
vacant site. If two adsorbates of
different species are nearest neighbors each other,
they react and form a product which desorbs immediately from the surface.
The system exhibits a first-order
transition from a phase saturated with one species
to another. Allowing desorptions of one species leads to
a continuous transition which belongs to the DP universality class
\cite{Albano,ZR}.

In both the monomer-monomer and the monomer-dimer
models, interactions between adsorbing particles
are ignored except for the actual
surface reaction which converts the reactants into the product.
Recently, an interacting monomer-monomer model \cite{Park,Zhuo}
has been introduced where particles of the same species have
{\it variable} repulsive interactions. When the interaction
strength is weak, the interaction model exhibits only
a first-order phase transition between two saturated
phases. At the critical value of the
interaction strength, a first-order line terminates at
a tricritical point beyond which two continuous-transition
lines appear. The two saturated phases are separated from
a reactive steady state by these two lines. These two
continuous transitions are shown to be in the DP
universality class again.

A common feature of these models is that they
exhibit, if  any, a continuous transition from a reactive phase into
an inactive phase of a {\it single} absorbing state. The resulting
critical behaviors are classified into the category of the so-called
``DP conjecture'' \cite{Jan,Grass82,Lai} which depicts that models
exhibiting a continuous-transition to a single absorbing state
generally belong to the universality class of the directed
percolation.

The universality class for models with a single
absorbing state is well established.  But few studies have
been made for models with more than one absorbing states.
Very recently
Jensen and Dickman \cite{Jen931,Jen932} have
extensively studied some nonequilibrium
lattice models with infinitely many absorbing states, the pair
contact process (PCP) and the reaction dimer (RD) model.
Both models have a continuous transition from a reactive phase
into an inactive phase with infinitely many absorbing states,
which is shown to
be, rather surprisingly, again in the DP universality class.

In contrast to the well-established DP universality class, only a few
models have appeared in literature which are known to be in the different
universality class from DP. Grassberger, Krause, and von der Twer
\cite{Grass84,Grass87} studied two models of probabilistic cellular
automata (PCA), namely model A and B. Both models exhibit a continuous
transition from an active phase into an absorbing phase, in which
the system is trapped into one of two (translationally) symmetric
states with particles and vacancies placed in the alternating sites.
But these models behave differently in the absorbing phase.
Once the system enters into one of the two
absorbing states, it remains in that state forever in model A,
but oscillates from one
state to the other in model B.
In spite of the discriminating behavior in the absorbing phase,
both models are shown to be in the same universality class but different
from DP.
The order-parameter exponent $\beta$ was obtained by
static Monte Carlo simulations, $\beta=0.6\pm 0.2$  \cite{Grass84}, but
later by
dynamic Monte Carlo simulations,
$\beta = 0.93\pm 0.12$ utilizing the hyperscaling relation
\cite{Grass87}.
The evolution rules for both models in common involve the processes
$X \rightarrow 3X$ and $2X \rightarrow 0$ that the number of particles
is conserved modulo 2. Grassberger has addressed in \cite{Grass87} that
this conservation law might be responsible for the non-DP behavior.

Another model which conserves the number of particles by modulo 2 can be
found in branching annihilating random walks with even number of
offsprings (BAW). Even though the BAW has a single absorbing state,
static Monte Carlo simulations \cite{Takayasu} show clearly non-DP
behavior with $\beta=0.7 \pm 0.1$ in one dimension. This value
seems to be consistent with that of the PCA, even though
heavy numerical errors obscure the fact.
The critical behavior of BAW with odd number of offsprings
is compatible with DP \cite{Jen933}.

All previous works \cite{Grass84,Grass87,Takayasu,Jen931,Jen932}
seem to imply as a whole that
more important in determining the universality
class of nonequilibrium phase transitions is not the number
of absorbing states but the mass conservation law in dynamics.
This shows a sharp contrast to the case of equilibrium phase transitions
where the number of ground states is relevant to the universality class.
The symmetry between absorbing states may be more important than the
number of absorbing states in determining the universality class.
In the PCP,
the infinitely many absorbing states are {\em not equivalent}
probabilistically.  Some absorbing states can be reached more easily
than other absorbing states by the PCP dynamics.
Therefore it is important to study a model with multiple {\em equivalent}
absorbing states.
The PCA models have two equivalent
absorbing states but also another conserving quantity, the
mass conservation
of modulo 2, which is claimed to be responsible for the non-DP behavior.
In this letter, we address a question whether a model with multiple
equivalent absorbing states may have non-DP behavior without any mass
conservation law.
As one of the simplest such models, we study the
monomer-dimer model with infinitely strong repulsive interactions
between the same species.

The interacting monomer-dimer model (IMD) is a generalization of the simple
monomer-dimer model, in which particles of the same species have
nearest-neighbor repulsive interactions. This is parameterized by
specifying that a monomer ($A$) can adsorb at a nearest-neighbor
site of an already-adsorbed monomer (restricted vacancy) at a rate
$r_Ak_A$ with $0 \leq r_A \leq 1$, where $k_A$ is an adsorption rate
of a monomer at a free vacant site with no adjacent monomer-occupied
sites. Similarly, a dimer ($B_2$) can adsorb at a pair of
restricted vacancies ($B$ in nearest-neighbor sites) at a rate
$r_Bk_B$ with $0 \leq r_B \leq 1$, where $k_B$ is an adsorption rate of
a dimer at a pair of free vacancies. There are no nearest-neighbor
restrictions in adsorbing particles of different species. The case
$r_A = r_B = 1$ corresponds to the ordinary noninteracting monomer-dimer
model which exhibits a first-order phase transition between two
saturated phases in one dimension. In the other limiting case
$r_A = r_B = 0$, there exists no fully saturated phase of monomers or
dimers. However, this does not mean that this model has no absorbing
states any more. In fact, there are two equivalent absorbing states in
this model. These states comprise of only the monomers at the
odd- or even-numbered lattice sites. There needs a pair of
adjacent vacancies for a dimer to adsorb, so a state with alternating
sites occupied by monomers can be identified with an absorbing state.

In this letter, we consider the IMD with
$r_A = r_B = 0$ only for simplicity.
General phase diagram of the IMD will appear elsewhere \cite{Mann}.
Then the system can be characterized by one parameter
$p = k_A/(k_A + k_B)$ of the monomer adsorption-attempt probability.
The dimer adsorption-attempt probability is given by
$q = 1 - p $.
The order parameter of the system is the concentration of dimers
$\bar{\rho}$ in the steady state, which vanishes algebraically as $p$
approaches the critical probability $p_c$ from below:
\begin{equation}
   \bar{\rho} \sim \left( p_c -p\right)^\beta,  \label{eq:beta}
\end{equation}
where $\beta$ is the order-parameter exponent. There are a
characteristic length
scale $\xi$ and time scale $\tau$ which diverge at criticality as
\begin{eqnarray}
   \xi  &\sim& |p_c -p|^{-\nu_\perp},\\
   \tau &\sim& |p_c -p|^{-\nu_\parallel},
\end{eqnarray}
where $\nu_\perp$ ($\nu_\parallel$) is a correlation length exponent
in the space (time) direction.
It is quite difficult to measure the order parameter accurately near
criticality by static Monte Carlo simulations due to the critical
slowing down. Moreover, there are strong finite-size effects near
criticality because of the diverging correlation length. In this letter,
we utilize the finite-size scaling (FSS) idea developed for nonequilibrium
phase transitions by Aukrust, Browne, and Webman \cite{Aukrust}.

Various ensemble-averaged quantities depend on system size through the
ratio of the system size and correlation length $L/\xi$. Thus we can
take the concentration of dimers near criticality as the following form:
\begin{equation}
   \bar{\rho} (p, L) \sim L^{-\beta/\nu_\perp}
                 f((p_c - p)L^{1/\nu_\perp})     \label{eq:rho}
\end{equation}
such that at $p_c$
\begin{equation}
   \bar{\rho} (p_c, L) \sim L^{-\beta/\nu_\perp}
\end{equation}
and
\begin{equation}
   f(x) \sim x^\beta ~~~\mbox{as}~~~ x \rightarrow \infty.
          \label{eq:asymp}
\end{equation}
In the supercritical region ($p < p_c$), the concentration $\bar{\rho}$
remains finite in the limit $L \rightarrow \infty$, but it should
vanish exponentially in the subcritical region ($p > p_c$).

For the time dependence of the order parameter
at criticality, one may assume a scaling form
\begin{equation}
   \bar{\rho} (L,t) \sim L^{-\beta/\nu_\perp}
   g(L t^{-\nu_\perp/\nu_\parallel}),  \label{eq:rho_t}
\end{equation}
so that for short time (or $L\rightarrow\infty$)
$\bar{\rho}$ has a power-law dependence on $t$ as
\begin{equation}
   \bar{\rho} \sim t^{-{\beta}/{\nu_\parallel}}.   \label{eq:rho_as}
\end{equation}
The characteristic time $\tau$ for a finite system is defined
as the elapsing time for a finite system to enter into
the (quasi-)steady state. An elementary FSS analysis finds
$\tau$ at $p_c$
\begin{equation}
   \tau (p_c, L) \sim L^{\nu_\parallel/\nu_\perp}. \label{eq:time}
\end{equation}

We run static Monte Carlo simulations. The initial
configuration is far from absorbing states with all sites vacated and
we use periodic boundary conditions. Then the system evolves along
the dynamical rules of our model. After one adsorption attempt on the
average per lattice site (one Monte Carlo step), the time is incremented
by one unit. The system reaches a quasi-steady state first and stays
for a reasonably long time before finally entering into an absorbing
state. We measure the concentration of dimers in the quasi-steady
state and average over some independent samples which have not yet
entered the absorbing states. The number of independent samples varies
from 50,000 for system size $L = 32$ to 300 for $L = 2,048$.
The number of time steps ranges from 500 to $2\times 10^5$ and, at least,
200 samples survive until the end of simulations.

At $p_c$, we expect the ratio of the concentrations of dimers for two
successive system sizes
$\bar{\rho}(L/2) / \bar{\rho}(L) = 2^{\beta/\nu_\perp}$,
ignoring corrections to scaling. This ratio converges to unity
for $p < p_c$ and diverges to infinity for $p > p_c$ in the limit
$L \rightarrow \infty$. We plot the logarithm
of this ratio divided by $\log 2$ as a function of $p$ for $L = 64$,
128, 256, 512, 1,024, and 2,048 in Fig.1. This plot shows strong
corrections to scaling. The crossing points between lines for two
successive sizes move to the right as the system size grows. In the
limit $L \rightarrow \infty$, we estimate the crossing points converge
to the point at $p_c \simeq 0.581(2)$ and $\beta/\nu_\perp \simeq 0.45(2)$.
The critical probability $p_c$ can be more accurately estimated from
dynamic Monte Carlo simulations \cite{COM}.
The value of the exponent ratio $\beta/\nu_\perp$ is clearly different
from the standard DP value of 0.2524(5) which is combined with
$\nu_\perp =1.0972(6)$ and $\beta = 0.2769(2)$.

In Fig.2, we have plotted $\bar{\rho}L^{\beta/\nu_\perp}$ against
$x = (1 - p/p_c) L^{1/\nu_\perp}$ in a double-logarithmic plot.
{}From Eqs.(\ref{eq:rho}) and (\ref{eq:asymp}), it follows that
for small $x$ the curve converges to a constant value, while
for large $x$ it should follow a straight line with slope $\beta$. We
find that the data for the various system sizes are best collapsed
to a curve with choices $p_c = 0.581(1)$, $\nu_\perp = 1.73(3)$, and
$\beta/\nu_\perp = 0.45(1)$, from which we get $\beta = 0.78(3)$. The
straight line in Fig.2 is a line with slope $0.78$, which
is in excellent accord with asymptotic behavior of $x$ in
Eq.(\ref{eq:asymp}).

By analyzing the decay characteristic of dimer concentrations
$\bar{\rho} (t,L)$,
we can determine the value of another correlation exponent
$\nu_\parallel$. First the finite-size behavior of
the characteristic time $\tau$ is investigated.
At $p=p_c \simeq 0.581$, we measure $\tau$
for the various
system sizes. The double-logarithmic
plot (see Fig.3) for the characteristic time $\tau$
versus the system size $L$
shows a straight line from which we obtain
the value $\nu_\parallel/\nu_\perp = 1.664(3)$. This
gives the value of $\nu_\parallel = 2.88(5)$, combined with
the value $\nu_\perp = 1.73(3)$. We also can check the value of
$\nu_\parallel$ independently by measuring the decay exponent of
$\bar{\rho}$ at $p=p_c$. From the
double-logarithmic plot for $\bar{\rho}(t)$
versus the time $t$ in Fig.4, we estimate
$\beta/\nu_\parallel=0.276(2)$, which is in excellent agreement with
the above results.
We summarize our results for the critical exponents as
$$
   \beta = 0.78(3),~~~\nu_\perp = 1.73(3),~~~\nu_\parallel = 2.88(5).
$$

In summary, we have numerically studied the interacting
monomer-dimer model with infinitely strong repulsive
nearest-neighbor interactions between the
same species in one dimension. The system exhibits a
continuous transition from a reactive phase into an inactive
phase with two equivalent absorbing states.
This model does not involve the dynamical processes
with the mass conservation of modulo 2. Nevertheless,
the values of the critical exponents are
clearly different from the DP values and seems to
be consistent with the values of the PCA and the BAW, in which the
the mass conservation of modulo 2 governs the dynamics.
Therefore we conclude that the symmetry between
absorbing states as well as the mass conservation law
are equally important in determining
the universality class for nonequilibrium phase
transitions. Up to date, it is not yet clear why
the model with two equivalent absorbing states and
the one with the mass conservation of modulo 2 seem to be in
the same universality class. Dynamical exponents and hyperscaling
relations of the IMD will be discussed elsewhere \cite{Mann}.

This work is supported in part by the Korean Science and
Engineering Foundation  through the grant No.931-0200-019-2
and also through the Center for Thermal and Statistical Physics
at Korea University.

{\center\Large Figure Captions}
\begin{description}
   \item[Fig.\ 1 :] Plots of $\log \left[ \bar{\rho}(L/2)/\bar{\rho}(L)
                  \right] / \log 2$ versus $p$. Open squares are for
      $L = 64$, solid squares for $L = 128$, open circles for $L = 256$,
      closed circles for $L = 512$, open triangles for $L = 1,024$, and
      closed triangles for $L = 2,048$.

   \item[Fig.\ 2 :] The double-logarithmic plot for the data of
     $\bar{\rho} L^{\beta/\nu_\perp}$ against
     $x = (1 - p/p_c)L^{1/\nu_\perp}$ for the various system size
     $L = 32 - 1,024$. With $p_c = 0.581$, $\nu_\perp = 1.73$,
     $\beta/\nu_\perp = 0.45$, the data are collapsed to a curve.
     The solid line is of slope 0.79 ($=\beta$).

   \item[Fig.\ 3 :] The characteristic time against the system
     size $L$ in a double-logarithmic plots for the various system
     size $L = 32 - 1,024$. The solid line is of slope 1.664
     $(=\nu_\parallel/\nu_\perp)$.

   \item[Fig.\ 4 :] The time-dependence of the concentration of
   dimers at $p_c = 0.581$. The slope of the curve
   gives the value of $\beta/\nu_\parallel = 0.276(2)$.

\end{description}


\begin{thebibliography}{99}
\bibitem{COM} The dynamic Monte Carlo method gives more systematic
        and accurate value for the critical point $p_c$.
        The obtained value is $p_c \simeq 0.581(1)$ which
          will be reported in details in \cite{Mann}.
\end{thebibliography}
\end{document}